\documentclass{article}
\usepackage{graphicx}  
\usepackage{amsmath}   
\usepackage[compress]{cite}
\usepackage{amssymb}   
\usepackage{bm} 
\usepackage{dcolumn}
\usepackage{color}
\usepackage{comment}
\usepackage{mathrsfs}
\usepackage{amsfonts}
\usepackage{varioref}
\usepackage{csquotes}
\usepackage{makecell}
\usepackage[normalem]{ulem}
\usepackage{float}
\usepackage[caption=false]{subfig}
\RequirePackage[colorlinks,citecolor=blue,urlcolor=magenta,linkcolor=blue]{hyperref}
\def\gr{general relativity}

\def\gr{general relativity}

\def \GB {Gauss-Bonnet}

\labelformat{chapter}{Chapter (#1)}
\labelformat{section}{Section (#1)}
\labelformat{subsection}{Section (#1)}
\labelformat{subsubsection}{Section (#1)}
\labelformat{subsubsubsection}{Section (#1)}
\labelformat{equation}{Eq.~(#1)}
\labelformat{figure}{Fig.~(#1)}
\labelformat{subfigure}{Fig.~(\thefigure#1)}
\labelformat{table}{Tab.~(#1)}
\labelformat{appendix}{Appendix #1}

\begin{document}
\title{Pure Gauss-Bonnet NUT Black Hole Solution: I}
\author{Sajal Mukherjee \footnote{mukherjee@asu.cas.cz}~$^{1,2}$, and Naresh Dadhich\footnote{nkd@iucaa.in}~$^{1}$ \\
$^{1}${\small{Inter-University Centre for Astronomy and Astrophysics, Post Bag 4, Pune-411007, India}}\\
$^{2}${\small Astronomical Institue of the Czech Academy of Sciences,}\\
{\small Bocni II 1401/1a, CZ-14100 Prague, Czech Republic}\\
}
\maketitle
\begin{abstract}
\noindent
We obtain an exact $\Lambda$-vacuum solution in the pure Gauss-Bonnet gravity with NUT charge in six dimension, with horizon having the product topology of $S^{(2)} \times S^{(2)}$. We discuss its horizon and singularity structure, and consequently arrive at parameter windows for its physical viability. It turns out that for the curvatures to remain function of $r$ alone for NUT black hole spacetime, horizon topology has to be product of $S^{(2)}$ spheres. This is true for Einstein as well as for pure Gauss-Bonnet gravity, and perhaps would hold good for higher order pure Lovelock as well. 
\end{abstract}
\section{Introduction}
With the recent developments in testing general relativity in various observational aspects \cite{Hulse:1974eb,Everitt:2011hp,Abbott:2016blz,Smoot:1992td,Perlmutter:1998np,Riess:1998cb}, the urge to explore alternative theories of gravity receives major scientific attention. While it is of particular interest to study these scenarios with observational implications  \cite{berti2015testing,Jana:2018djs}, there also exists sufficient motivation to obtain new black hole (BH) solutions in these theories of gravity \cite{Dadhich:2015nua}. It is a general belief that deviation from \gr~is expected to have higher curvature corrections terms \cite{Stelle:1977ry,Brandenberger:1993ef}. It is therefore pertinent to study these theories in detail. Two of the widely known higher curvature theories of gravity include $f(R)$ \cite{Sotiriou:2008rp}, and Lovelock theory of gravity \cite{Lovelock:1971yv}. The $f(R)$ theory of gravity was initially introduced to explain the accelerating expansion of the Universe without invoking any nontrivial matter components such as dark energy \cite{Nojiri:2010wj,DeFelice:2010aj}. Since then, $f(R)$ theory has been a subject of immense activity, and studied extensively in Refs. \cite{Chakraborty:2016ydo,Datta:2019npq,Vainio:2016qas}. On the other hand, Lovelock gravity was introduced by David Lovelock in 1971, and stood out as an excellent and natural higher dimensional generalization of general relativity. Given that Lovelock theory of gravity being our primary focus, we briefly recount some of the prime aspects of it in the following paragraph.

The striking and remarkable feature of the Lovelock theory is that the equation retains the second order character despite the action being polynomial in Riemann curvature. This is in contrast to all other alternative theories where the equation of motion involves physically undesirable higher order derivatives. It is therefore free of ghosts \cite{Zwiebach:1985uq,Deruelle:1989fj}. Note that the Einstein's general relativity (GR) is linear order, $N=1$ in the Riemann curvature in action, the quadratic $N=2$ is the well known Gauss-Bonnet (GB) \cite{Boulware:1985wk,Bajardi:2019zzs}, and so on. Each order comes with a dimensionful coupling constant. The particular order $N$ term makes non-zero contribution to the equation of motion only in dimensions, $D\geq 2N+1$. It is therefore quintessentially higher dimensional generalization of GR.

Pure Lovelock theory is specified by the property that the action has only one chosen $N$th order term without sum over the lower orders; i.e. pure GB action will have only the quadratic term with no Einstein term. Pure Lovelock gravity is characterized by some interesting and distinguishing properties: (a) in the critical odd $D=2N+1$, it is kinematic \cite{Dadhich:2012cv,Camanho:2015hea}, in the sense that $N$th order Lovelock Riemann is entirely given in terms of the corresponding Ricci. It also hints that vacuum solution being trivial, in all critical odd dimensions $D=2N+1$, for example., $D=5$ for pure Gauss-Bonnet. Therefore, non-trivial vacuum solutions in pure Gauss-Bonnet gravity would only exist in dimension $D \geq 6$.  (b) existence of bound orbits around a static object in higher dimensions \cite{Dadhich:2013moa} and (c) stability of static BH \cite{Gannouji:2019gnb} (for an insightful overview, we refer our readers Ref. \cite{Dadhich:2015nua}).

In this paper, we choose to obtain a pure GB static BH solution with the NUT charge in higher dimension. Note that static solutions in pure Lovelock with vanishing NUT charge are already studied in \cite{Cai:2006pq,Garraffo:2008hu,Dadhich:2012ma}. Here, we will explore what additional features the NUT charge will bring into the BH solution. NUT was a very strange and unusual solution of Einstein's equation \cite{Newman:1963yy}. There has been much discussion on its physical understanding and interpretation, see for instance an extensive account in \cite{LyndenBell:1996xj}. Apart from physical interpretation, several recent studies \cite{Chakraborty:2017nfu,Chakraborty:2019rna} also involve observational implications and astrophysical imprints of NUT charge. In addition to other interesting properties of NUT parameter, it is endowed with a remarkable features of duality. It is shown by Turakulov and Dadhich in Refs. \cite{LyndenBell:1996xj, Turakulov:2001jc}, that with the presence of both rotation and NUT parameters, the spacetime is invariant under a duality  transformation where mass and NUT charge and radial and angular coordinates are interchanged. This indicates that the NUT parameter is associated with a gravomagnetic field and can be considered as a gravomagnetic charge \cite{Turakulov:2001bm}. It could as well be looked upon as a rotating gravitational dyon. In fact the Kerr-NUT solution is the most general axially symmetric solution that admits separability of Hamilton-Jacobi and Klein-Gordon equations \cite{Dadhich:2001sz}. Alongside these studies, the geodesic trajectories and orbital dynamics, thermodynamics properties and the Lense-Thirring effect etc., have also been investigated  \cite{Mukherjee:2018dmm}.

The NUT parameter comes as a very natural extension of the Kerr-Newman family of BHs when the asymptotic flatness condition is relaxed \cite{LyndenBell:1996xj}. In dimensions greater than four, all NUT BH solutions have product of $S^{(2)}$ spheres topology like $S^{(2)} \times S^{(2)}$ in six dimensions. In Ref. \cite{Pons:2014oya}, higher dimensional BH solutions with the product topology are obtained in Einstein and Einstein GB gravity. Here, we wish to obtain a pure GB black hole solution with product topology in the presence of NUT charge. In this Part I of the investigation, we shall confine to pure GB NUT BH while its Maxwell charge generalization would be taken up separately in Part II \cite{sajal_mukherjee}. We also refer readers to Ref. \cite{Mukherjee:2020lld}, where some of the results of this article are highlighted in a short and concise manner. The paper is organized as follows: In \ref{sec:Black_GB_EN}, we address the BH solutions in both Einstein and pure Gauss-Bonnet gravity for different horizon topology. We also set up the necessary field equations for obtaining both Einstein and pure GB vacuum solutions. In \ref{sec:topology} we demonstrate how the horizon topology affects the higher dimension NUT BH solutions. Following this, in \ref{sec:Gauss_NUT}, we obtain a new  pure \GB~NUT BH solution and discuss its various properties and spacetime structure in \ref{sec:structure}. Finally, we close the article with a brief discussion in \ref{sec:discussion}.

\textit{Notation and convention:} Throughout the paper, we use \enquote*{D} as dimension of spacetime, \enquote*{prime} as a differentiation with respect to the radial coordinate $r$, and the \enquote*{bracket} gives a projected quantity, ${X}^{(\mu) (\nu)}=e^{(\mu)}_{~\alpha}e^{(\nu)}_{~\beta} X^{\alpha \beta}$, on a tetrad frame. The Greek letters $\mu$, $\nu$ run for both temporal and spatial components, while Latin letters $i$, $j$, only for spatial components. Besides, we shall adopt the metric signature:  $(-, +, +, ...)$ and set the fundamental constants as $c = 1 = G$.
\section{Warming up: static BH solutions for different horizon topology} \label{sec:Black_GB_EN}
In this section, we introduce static BH solution in pure Gauss-Bonnet gravity in six dimension without the NUT parameter, and  set up the pure GB field equations for later use. We discuss BHs with spherical and product topologies and consider their  horizon and spacetime structure properties for future reference in relation to pure GB NUT BH solution.
\subsection{Gauss-Bonnet action and field equations} \label{sec:Preli}
The gravitational action for the pure Gauss-Bonnet gravity in D dimension is given as
\begin{equation}
\mathcal{S}= \int d^Dx \sqrt{-g}\bigl(\mathcal{L}_{\rm GB}+\mathcal{L}_{\rm m}+\Lambda \bigr),
\end{equation}
where $\mathcal{L}_{\rm m}$ is the matter Lagrangian, and the GB Lagrangian, $\mathcal{L}_{\rm GB}$ is given by
\begin{equation}
\mathcal{L}_{\rm GB}=R^2-4R^{ab}R_{ab}+R^{abcd}R_{abcd}.
\label{eq:LGB}
\end{equation}
By varying the action with respect to the metric, we obtain the field equations (GB coupling constant $\alpha_2=1$) in the usual notation,
\begin{equation}
H_{ab}=\left(2 J_{ab}-\dfrac{1}{2}g_{ab} \mathcal{L}_{\rm GB} \right)=-\Lambda g_{ab}+T_{ab}.
\label{eq:field_equation}
\end{equation}
In the above equation $H_{ab}$ and $J_{ab}$ play the role analogous to Einstein and Ricci tensor in Einstein's gravity \cite{Padmanabhan:2013xyr} and the latter is defined as
\begin{equation}
J_{ab}=RR_{ab}-2R^{c}_{~a}R_{bc}-2R^{cd}R_{acbd}+R_{a}^{~mnl}R_{bmnl}.
\label{eq:Jab}
\end{equation}
These equations are solved for vacuum by setting $T_{ab}=0$ which we discuss next.
\subsection{The metric structure} \label{sec:metric}
In higher dimensions, BH horizon can have different topologies. For instance in six dimension, there are following possible three choices: the  spherical $S^{(4)}$, product of two 2-spheres, $S^{(2)} \times S^{(2)}$, and product of 3-sphere and a line element, $S^{(1)} \times S^{(3)}$ \cite{Hervik:2019gly,Hervik:2020zvn}. Out of these options, we would only consider the first two cases in this investigation. We shall first begin with the spherical and then take up the product topology.
\subsubsection{Spherical topology:}
In the case of a spherical topology, which is given by a 4-sphere in six-dimension, the metric ansatz takes the following form in $(t,r,\theta_1,\theta_2,\theta_3,\phi)$ coordinates:
\begin{equation}
ds^2=-A(r) dt^2+{B(r)}dr^2+{r^2} d\Omega^2_4,
\label{eq:metric_ansatz_eg_01}
\end{equation}
where $d\Omega^2_4$ can be expanded as
\begin{equation}
d\Omega^2_4=d\theta^2_3+\sin^2\theta_3~d\theta^2_2+ \sin^2\theta_2 \sin^2\theta_3 d\theta^2_1+\sin^2\theta_1 \sin^2\theta_2 \sin^2\theta_3 d\phi^2.
\label{eq:spherical_topology}
\end{equation}
We now employ the above metric, and obtain the following expressions for $H^{0}_{0}$ and $H^{1}_{1}$
\begin{eqnarray}
 & & H^{0}_{0}=-\dfrac{12\Big(1-B(r)\Big)}{r^4 B(r)^2}\Big\{1-B(r)-2rB^{\prime}(r)/B(r)\Big\}=-\Lambda,\nonumber \\
 & & H^{1}_{1}=-\dfrac{12\Big(1-B(r)\Big)}{r^4 B(r)^2}\Big\{1-B(r)+2 r A^{\prime}(r)/A(r)\Big\}=-\Lambda.
\end{eqnarray}
By computing $H^{0}_{0}-H^{1}_{1}=0$,
we arrive at for $B(r) \neq 1$,
\begin{equation}
 A(r)B(r) = \text{const}.
\end{equation}
Now asymptotically both $A$ and $B$ should tend to unity and therefore, $A(r)B(r)=1$.  Finally, the field equations become:
\begin{eqnarray}
H^{0}_{0} &=& H^{1}_{1} = \dfrac{12}{r^4} \left\{2 r A^{\prime}(r)+2A(r)-\bigl[A(r)\bigr]^2-2 r A(r) A^{\prime}(r)-1\right\}=-\Lambda,  \\
H^{2}_{2} &= & H^{3}_{3}=H^{4}_{4}=H^{5}_{5}=-\dfrac{6}{r^3}\left\{2 \bigl[A(r)-1 \bigr]A^{\prime}(r)+r \bigl[A^{\prime}(r)\bigr]^2+r \bigl[A(r)-1\bigr]A^{\prime \prime}(r) \right\}=-\Lambda. \nonumber \\
\end{eqnarray}
From the above equations, it clearly follows that $H^{1}_{1}$ and $H^{2}_{2}$ are no longer independent, and are in fact related by the following relation,
\begin{equation}
H^{2}_{2}=\dfrac{r}{4}\dfrac{d H^{1}_{1}}{dr}+H^{1}_{1},
\label{eq:H22_H11_New}
\end{equation}
This clearly shows that one needs to solve only the first order equation, $H^{0}_{0}=-\Lambda$ for $A(r)$, rest of the equations are then automatically satisfied. It solves to give the solution,
\begin{equation}
A(r)=1-\sqrt{\dfrac{M}{r}+\dfrac{\Lambda r^4}{60}},
\label{eq:S4_GB_Cosmo}
\end{equation}
where, $M$ is the integration constant indicating mass of the BH. A general solution with arbitrary order and dimension in pure Lovelock gravity is obtained in Ref. \cite{Dadhich:2012ma}.
\subsubsection{Non-spherical topology:}  \label{sec:non_sph_gb}
In this case, we start with the following metric ansatz in $(t,r,\theta_1,\phi_1,\theta_2,\phi_2)$ coordinates \cite{Dehghani:2005zm}:
\begin{equation}
ds^2=-A(r) dt^2+{B(r)}dr^2+{r^2} \left\{d\Omega^2_{1} + d \Omega^2_{2} \right\},
\label{eq:metric_ansatz_eg_gb_02}
\end{equation}
where
\begin{equation}
d \Omega^2_{1}=d\theta_1^2+\sin^2\theta_1 d\phi_1^2, \quad d \Omega^2_{2}=d\theta_2^2+\sin^2\theta_2 d\phi_2^2.
\label{eq:product_surface}
\end{equation}
For the above metric, we obtain the following expressions of $H^0_0$ and $H^1_1$:
\begin{eqnarray}
H^0_0=-\dfrac{12 \left \{B(r)-2 B^2(r)+3 B^3(r)-2 r B^{\prime}(r)+2 r B(r)B^{\prime}(r)\right \}}{r^4 B^3(r)}, \nonumber \\
H^1_1=\dfrac{12\left\{A(r)(-1+2 B(r)-3 B^2(r))+2 r (-1+B(r))A^{\prime}(r)\right\}}{r^4 A(r)B^2(r)}.
\end{eqnarray}
Again $H^0_0-H^1_1=0$, as before would imply $A(r)B(r) = 1$. With this we write
\begin{eqnarray}
H^{0}_{0} &=& H^{1}_{1} = -\dfrac{4}{r^4} \left\{1+3 (A(r))^2-2 r A^{\prime}(r)+A(r)\bigl[-2+6 r A^{\prime}(r)\bigr]\right\}=-\Lambda, \label{eq:H00_Sph_GB} \\
H^{2}_{2} &= & H^{3}_{3}=H^{4}_{4}=H^{5}_{5}=\dfrac{1}{r^3}\left\{\bigl[4-12 A(r)\bigr]A^{\prime}(r)-6 r \bigl[A(r) \bigr]^2+2 r \bigl[1-3 A(r) \bigr]A^{\prime \prime}(r) \right\}=-\Lambda. \nonumber \\
\end{eqnarray}
Interestingly, for this product topology too, $H^{2}_{2}$ and $H^{1}_{1}$ are related by the same relation \ref{eq:H22_H11_New}, and again we solve the first order equation to get the solution \cite{Dadhich:2015nua}:
\begin{equation}
A(r)=\dfrac{1}{3} \left\{1-\sqrt{\dfrac{M}{r}-2+\dfrac{3 \Lambda r^4}{20}}\right\}.
\label{eq:EH_GB_S2 X S2}
\end{equation}
The factor $1/3$ above appears as a consequence of product topology. For an arbitrary dimension $d$, this pre-factor is given by ${(d-4)}/{(2d-6)}$. This factor could as well be shifted to the product of spheres, i.e., by writing, $(1/3) r^2d\Omega^2_{1,2}$ instead of $r^2 \Omega^2_{1,2}$ in the metric. With this shift of the factor, and we write,
\begin{equation}
A(r)= \left\{1-\sqrt{\dfrac{M}{r}-2+\dfrac{ \Lambda r^4}{60}}\right\},
\label{eq:EH_GB_S2 X S2_New}
\end{equation}
where $M$ and $\Lambda$ have been appropriately rescaled. However, the factor of $-2$ that appears inside the root is also due to product topology which is absent in \ref{eq:S4_GB_Cosmo} for the spherical case. In GB gravity with topology, $S^{d_0} \times S^{d_0}$, this would read as $- d_0/(d_0-1)^2(2d_0-3)$ which for $d_0=2$ would be $-2$ \cite{Dadhich:2012pda,Pons:2014oya}.

Product topology produces topological defect as solid angle deficit and to counter its effect, the prefactor like $1/3$ is needed. That is what happens in linear order Einstein gravity but in the quadratic GB, this is not enough and an additional factor $-2$ is needed under the square root. There is a very interesting and insightful interplay between these two topological factors, $p$ and $q$ where the former stands for $1/3$ and the latter for $-2$, in terms of solid angle deficit as elaborated in \cite{Dadhich:2015nua}.
%
\subsection{Spacetime singularity} \label{sec:singularity}
The topology of the horizon can impart nontrivial effect in shaping the singularity structure of the spacetime. Recalling from Ref. \cite{Pons:2014oya} the solution given by \ref{eq:EH_GB_S2 X S2_New}, there occurs, apart from the central singularity at $r=0$, an additional non-central singularity as could be seen from the following expression for Ricci scalar,
\begin{equation}
R = \dfrac{1}{4r^4}\dfrac{1}{(M/r-2+r^4 \Lambda/60)^{3/2}}\Bigl\{70M^2+6Mr(r^4 \Lambda-56)+\dfrac{r^2}{15}(r^4 \Lambda-144)(r^4 \Lambda-40)\Bigr\}.
\label{eq:spacetime_sing}
\end{equation}
The non-central singularity is given by $X=-2 + M/r + \Lambda r^4/60 =0$. Note that for the solution to be real in \ref{eq:EH_GB_S2 X S2_New}, $X \geq 0$ is required, while the horizon is defined by $X=1$. However, what it indicates is that $X$ would always appear in the denominator for the Kretschmann like scalar, in particular GB Kretschmann scalar, constructed from the Gauss-Bonnet analog of Riemann tensor $\mathcal{R}_{ijkl}$ \cite{Dadhich:2008df}. Therefore, for the validity of the solution, non-central singularity has to lie inside the horizon. A detailed analysis of the parameter range required for physical validity of the solution has been comprehensively discussed in \cite{Pons:2014oya}.
\subsection{The event horizon}
\begin{figure}
\centering
\includegraphics[scale=.4]{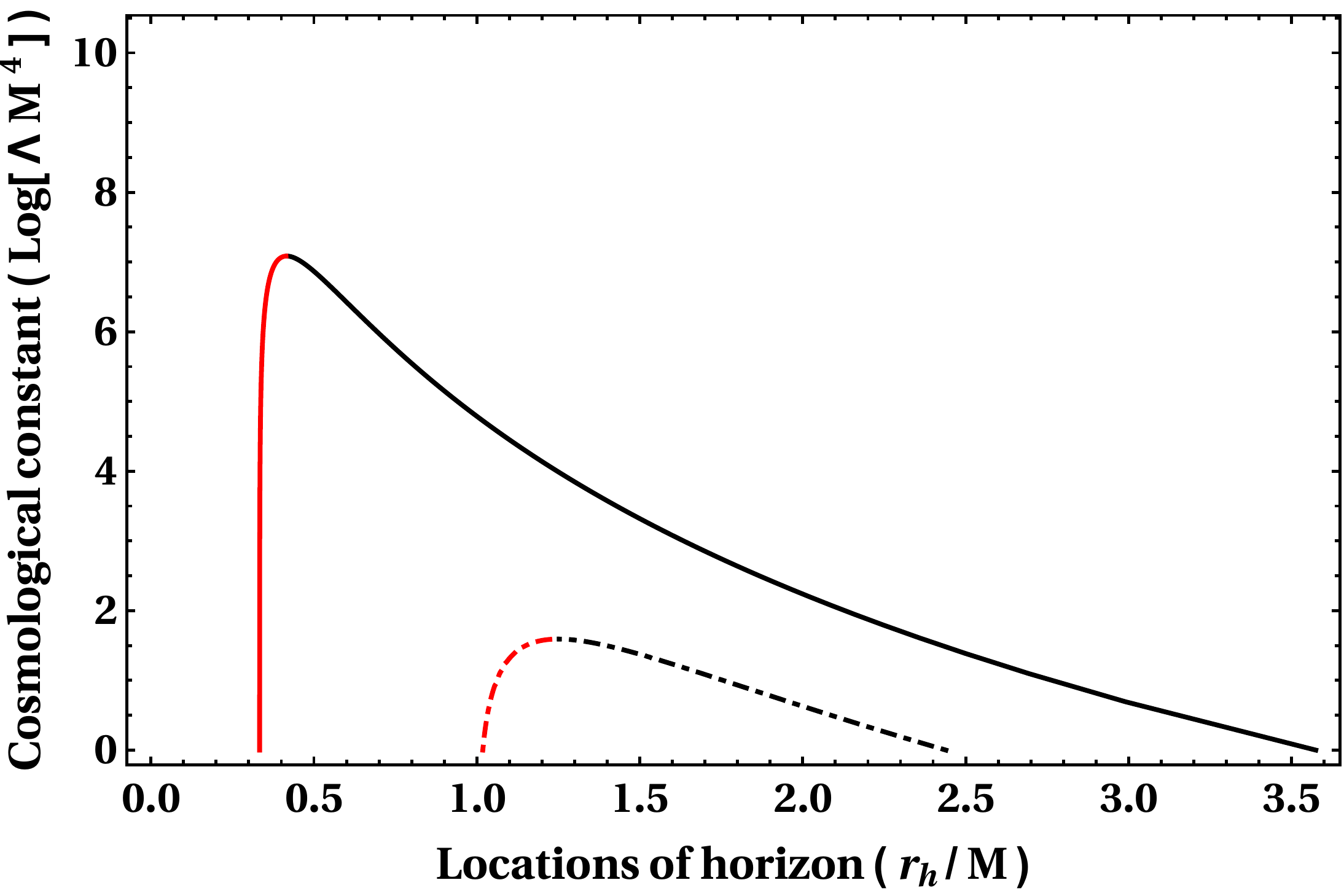}
\caption{The location of  event horizon (red curve) and cosmological horizon (black curve) for different topologies ---  solid and dashed curve respectively representing $S^{(2)} \times S^{(2)}$ and $S^{(4)}$. Interestingly, the cut-off on $\Lambda$ is larger for the spherical case than the product topology, which is due to the additional factor $-2$ in the square root. For appropriate scaling of the plots, we have used $\log({\Lambda}M^4)$ instead of ${\Lambda M^4}$.}
\label{Fig_01}
\end{figure}

Before closing this section, we consider a simple example to demonstrate the difference in location of event horizon in $S^{(2)} \times S^{(2)}$ and $S^{(4)}$ topology in pure Gauss-Bonnet gravity. In these two cases, $A$ is given in  \ref{eq:S4_GB_Cosmo} for $S^{(4)}$ and in \ref{eq:EH_GB_S2 X S2_New} for $S^{(2)} \times S^{(2)}$ topology. The horizon is defined by $A=0$ which is a fifth order equation having two positive roots giving event and cosmological horizons. \ref{Fig_01} shows horizon curves for the two cases where red and black respectively indicate event and cosmological horizons, and solid and dashed lines refer to $S^{(2)} \times S^{(2)}$ and $S^{(4)}$ topologies. In each of the cases, there exists a maximum value of the cosmological constant beyond which there exists no horizon. For $S^{(4)}$ it is
\begin{equation}
0<\log({\Lambda} M^4) \lessapprox 1.5922,
\end{equation}
while for $S^{(2)}\times S^{(2)}$,
\begin{equation}
0<\log({\Lambda} M^4)\lessapprox 7.0853.
\end{equation}
\noindent
With this, we finish our discussion on the higher dimensional BH solutions in spherical and product topology.  In the upcoming discussions, we would include the NUT charge explicitly, and study its consequences.

\section{The interplay between topology and NUT parameter} \label{sec:topology}
The presence of NUT charge in higher dimensional BH solutions introduces a topological impediment, which we have found to be true for both Einstein and Gauss-Bonnet gravity. Due to this limitation, there exists \textit{no} higher dimensional NUT BH with spherical topology, however, they do exist with non-spherical product topology  \cite{Flores-Alfonso:2018jra,Awad:2000gg,Corral:2019leh}. But there is hardly any discussion about why spherical topology is not allowed? This is the question we wish to address here.

We know that horizon of four dimensional NUT BH has $S^{(2)}$  topology and its Riemann curvature is free of the angle coordinate and depends on $r$ alone. What happens when we go to higher dimensions? It turns out that this character of Riemann curvature, that it is function of $r$ alone, could only be maintained for horizon topology being product of $S^{(2)}$ spheres like $S^{(2)} \times S^{(2)}$ for $D=6$. In the following, we show with an explicit computation that whenever it is not product of $S^{(2)}$ spheres Riemann is not a function of $r$ only. We would however compute the Riemann component in the orthonormal frame so that it has physical meaning and free from coordinate dependent pitfalls. In this context whatever is true for Riemann tensor would also be true for Lovelock Riemann tensor because it is a homogeneous polynomial in Riemann \cite{Camanho:2015hea}.

For illustration, let us begin with the metric,
\begin{equation}
ds^2=-\mathcal{F}(r)\bigl(dt+2 l \cos\theta_1 d\phi\bigr)^2+\dfrac{1}{\mathcal{F}(r)} dr^2+ (r^2+l^2)  d\Omega^2,
\label{eq:metric_ansatz}
\end{equation}
where, $l$ being the NUT charge, and $d\Omega^2$ can either be a 4-sphere (as in \ref{eq:spherical_topology}), or the product $S^{(2)} \times S^{(2)}$ (as in \ref{eq:product_surface}), and $\mathcal{F}(r)$ is arbitrary. For finding the solution one has to solve for the function $\mathcal{F}(r)$.

We shall now compute a Riemann component, $R^{(0)}_{~(2)(0)(2)}$ for the above metric in the orthonormal frame which is given by:
\begin{equation}
R^{(0)}_{~(2)(0)(2)}=-\dfrac{r {\mathcal{F}}^{\prime}(r)}{2(r^2+l^2)}-\dfrac{ l^2 \mathcal{F}(r)(\csc\theta_2)^4 (\csc\theta_3)^4}{(r^2+l^2)^2},
\end{equation}
for $S^{(4)}$ topology. This clearly depends upon the angle coordinates so long as $l\neq0$.
On the other hand for the product $S^{(2)} \times S^{(2)}$ topology given in \ref{eq:product_surface}, it is
\begin{equation}
{R}^{(0)}_{~(2)(0)(2)}=-\dfrac{r {\mathcal{F}}^{\prime}(r)}{2(r^2+l^2)}-\dfrac{ l^2 \mathcal{F}(r)}{(r^2+l^2)^2},
\label{Eq:Riemann_Sph}
\end{equation}
which is clearly independent of angular coordinates. It can be easily verified that whatever is true for this one component is true for all other Riemann components.

Moreover, to strengthen our claim that the radial dependence may be responsible for the topological limitation in higher dimensional NUT solution, we also consider eight dimensional case with the topology $S^{(6)}$, product of three $S^{(2)}$ and product of two $S^{(3)}$. It turns out that radial dependence is preserved only for product of $S^{(2)}$ case and none other. We obtain in particular,
\begin{eqnarray}
\left. {R}^{(0)}_{~(2)(0)(2)} \right \vert_{S^{(6)}}&=& \left. {R}^{(0)}_{~(2)(0)(2)}\right \vert_{S^{(2)}\times S^{(2)}\times S^{(2)}}+\dfrac{l^2 \mathcal{F}(r)\left[1-(\csc\theta_2)^4 (\csc\theta_3)^4 (\csc\theta_4)^4 (\csc\theta_5)^4\right]}{(r^2+l^2)^2}.
\end{eqnarray}
Like \ref{Eq:Riemann_Sph} the first term on the right is a function of $r$ alone. We have also verified for the case of product of two ($S^{(3)}$) that the angle dependence cannot be avoided. It is only product $S^{(2)}$ spheres that alone ensures Riemann to be function of $r$ alone.

Since Riemann tensor is the basic element from which gravitational theory emerges, Einstein or Lovelock, therefore the above result of $S^{(2)}$ product topology and radial dependence of Riemann tensor for NUT spacetime would hold good in general. 
\section{Pure GB NUT Black hole solution} \label{sec:Gauss_NUT}
%
In this section, we solve pure GB $\Lambda$-vacuum equation in the presence of NUT parameter. We obtain a new exact solution describing a pure NUT BH with $\Lambda$ in six dimension with product topology $S^{(2)} \times S^{(2)}$. Keeping in mind the above discussion on product topology, and its prefactor, $(1/3)$ in \ref{eq:EH_GB_S2 X S2_New}, we write the metric for $S^{(2)} \times S^{(2)}$ topology as follows:
\begin{equation}
ds^2=-\dfrac{\Delta}{\rho^2}\Bigl\{dt + P_1 d\phi_1 +P_2d\phi_2 \Bigr\}^2+\dfrac{\rho^2}{\Delta}dr^2+\dfrac{\rho^2}{3} \left\{d\theta_1^2+\sin^2\theta_1 d\phi^2_1+d\theta_2^2+\sin^2\theta_2 d\phi^2_2 \right\},
\label{eq:metric_S2_S2_SNUT}
\end{equation}
where $l$ being the NUT parameter, $\rho^2=r^2+l^2$, $\Delta=\rho^2 \left[1-f(r)\right]$, $P_1={2 l \cos\theta_1/3}$ and $P_2={2 l \cos\theta_2/3}$. Note that in the above metric, we have already incorporated the condition,  $H^{(0)}_{(0)}=H^{(1)}_{(1)}$, and product topology prefactor as in \ref{sec:non_sph_gb} and Ref. \cite{Dadhich:2015nua}. The $H_{ab}$ components then take the form:
\begin{eqnarray}
H^{(0)}_{(0)}=H^{(1)}_{(1)}=\dfrac{12}{(r^2+l^2)^3}\Big\{-(r^2+l^2)(f(r))^2-2\Big(r^2+2rl^2 f^{\prime}(r)+3l^2\Big)+2 f(r)\Big(r(l^2-r^2)f^{\prime}(r)+2l^2\Big)\Big\}, \nonumber  \\
H^{(2)}_{(2)}=H^{(3)}_{(3)}=H^{(4)}_{(4)}=H^{(5)}_{(5)}=\dfrac{6}{(r^2+l^2)^3}\Big\{4l^2rf^{\prime}(r)+(l^4-r^4)(f^{\prime}(r))^2-2l^2\Big((r^2+l^2)f^{\prime \prime}(r)-2\Big)+\nonumber \\
+ f(r)\Big(-4l^2-2r(r^2+3l^2)f^{\prime}(r)+(l^4-r^4)f^{\prime \prime}(r)\Big)\Big\}.\nonumber
\\
\end{eqnarray}
Interestingly, in this case too, we obtain a relation between $H^{(2)}_{(2)}$ and $H^{(1)}_{(1)}$ written as:
\begin{equation}
H^{(2)}_{(2)}=\dfrac{r^2+l^2}{4r}\dfrac{dH^{(1)}_{(1)}}{dr}+H^{(1)}_{(1)}, \label{eq:H22_H11_l}
\end{equation}
which reduces to \ref{eq:H22_H11_New} once we set $l=0$. By solving the first order differential equation $H^{(0)}_{(0)} = - \Lambda$ as given in \ref{eq:H00_Sph_GB} which gets readily solved to give
\begin{eqnarray}
f(r) &=& \dfrac{1}{90(l^4-r^4)}\Bigl[180 l^2 (r^2+l^2)-\sqrt{15}\Bigl\{9 (r^2+l^2)^2\Bigl[5 l^8 \Lambda-20 l^6 r^2 \Lambda +10 l^4 (-12 +r^4 \Lambda)\nonumber \\
& & + r^3 \bigl(60 M-120 r+r^5 \Lambda \bigr)+4 l^2 r \bigl(-15 M+120 r+r^5 \Lambda\bigr) \Bigr] \Bigr\}^{1/2}\Bigr],
\label{eq:f(r)_charged}
\end{eqnarray}
where $M$ is the integration constant identified with BH mass. This is a new solution describing a pure GB NUT BH in six dimensions with product topology $S^{(2)} \times S^{(2)}$. For the further discussion, we rewrite the above equation in the following form,
\begin{equation}
f(r)=\dfrac{1}{l^2-r^2}\Bigl[{2 l^2-\Big\{h(r)\Big\}^{1/2}}\Bigr],
\label{eq:f(r)}
\end{equation}
where
\begin{eqnarray}
h(r)=\dfrac{1}{60}\Bigl\{5 l^8 \Lambda-20 l^6 r^2 \Lambda +10 l^4 (r^4 \Lambda -12)+4 l^2 r (120 r+r^5 \Lambda-15 M)+r^3(-120r+r^5 \Lambda+60M)\Bigr\}.
\label{C_1_0_C_2_0}
\end{eqnarray}
\noindent
Let's now consider some of interesting limiting cases.
\begin{enumerate}
\item In the limit of vanishing NUT charge, i.e., ${l}=0$, we retrieve the known solution,
\begin{equation}
f({r})\bigg|_{{l}=0}=\sqrt{\dfrac{M}{{r}}-2+\dfrac{{\Lambda} {r}^4}{60}},
\label{eq:Limit_case}
\end{equation}
as given in Eq. (16) of \cite{Dadhich:2015nua}.
\item  In the asymptotic limit ${r \to \infty}$, we have
\begin{equation}
f({r})\bigg|_{{r} \rightarrow \infty} = \left \{{\Lambda} \Big(\frac{{r}^4}{60}\Big)-2\right\}^{1/2}.
\label{eq:f(r)_S2XS2_r_infinity}
\end{equation}
As in \ref{eq:Limit_case} above, this limit exists only when $\Lambda > 0$ is present, i.e., presence of positive $\Lambda$ is therefore essential for pure GB BH with product topology.
\item \ref{eq:f(r)} appears to diverge when $r=l$. However that is not really the case when we take the limit $r\to l$ properly and it actually reads as
\begin{equation}
f({r})\bigg|_{{r} = l} = 1+\dfrac{M}{4l}+\dfrac{\Lambda l^4 }{15},
\label{eq:f(r)_r=l}
\end{equation}
which is finite so long as $l \neq 0$.
\end{enumerate}
With this, we finish the discussions on limiting cases and shall now take up  discussion of singularity and horizon.
\subsection{Singularity}
As we have seen earlier in \ref{eq:EH_GB_S2 X S2_New} that it is the term under the root, the discriminant is required to be non-negative  for reality of the solution and when it vanishes it makes \ref{eq:spacetime_sing} singular. In the present spacetime under consideration, $h(r) = 0$ marks the non-central singularity, and hence $h(r) > 0$ is required for physical viability of the solution.

Note that whenever NUT parameter is present there occurs no central singularity, spacetime is singular only at $h(r)=0$, which indicates non-central singularity. For physical viability of the solution, this has to be covered by an event horizon so that it is naked. The polynomial equation $h(r) = 0$ given in \ref{C_1_0_C_2_0} can be rewritten in descending order as:
\begin{eqnarray}
h(r)=\Bigl[\dfrac{\Lambda}{60} r^8+\dfrac{\Lambda}{15}l^2 r^6+\dfrac{1}{6}\left(\lambda -12\right)r^4+M r^3-\dfrac{l^2 r^2}{3}\left(\lambda -24\right)-M l^2 r+\nonumber \\
\dfrac{l^4}{12}\left(\lambda -24\right)\Bigr]=0 ,
\label{C_1_0_C_2_0_NEW}
\end{eqnarray}
where we have defined the dimensionless parameter, $\lambda = \Lambda l^4$ which determines number of roots the above equation can have. For instance, when $\lambda < 12$, it can have three positive roots while it can have only one for $\lambda = 12$.

%
\subsection{Horizon structure}
The location of the horizons $(r_{\rm h})$ is given by $f(r_{\rm h})=1$, and from \ref{eq:f(r)}, this becomes, $h(r_{\rm h})=(r_{\rm h}^2+l^2)^2$. By using the expression of $h(r)$ given in \ref{C_1_0_C_2_0}, we gather
\begin{equation}
{h}(r_{\rm h})-(r_{\rm h}^2+l^2)^2=-\dfrac{l^2-r_{\rm h}^2}{60} \mathcal{X}(r)=0.
\label{eq:h_tilder_h}
\end{equation}
In view of \ref{eq:f(r)_r=l}, $r_{\rm h}^2 \neq l^2$, hence horizon would be given by
\begin{equation}
\mathcal{X}(r)=r_{\rm h}^6 \Lambda+5 \Lambda l^2 r_{\rm h}^4+15 r_{\rm h}^2(\lambda -12)+60 M r_{\rm h}-5 l^2(\lambda -36) = 0 ,
\label{eq:Xr}
\end{equation}
where $\lambda = \Lambda l^4$ as defined earlier.
The roots analysis of this polynomial could be carried out similarly and it is readily evident that it has no positive roots, i.e. no horizons, in the parameter window: $12 \leq \lambda \leq 36$. For horizon to occur, $\lambda$ has to be outside this range. Further  $h(r)>0$ has also to be ensured for physical viability.

To highlight an interesting property of the above equation, which will be relevant in the upcoming discussion, we consider large and small $r$ limit of the above equation. This is useful for  deciding whether the horizon is event or cosmological. For large $r$ limit, we have
\begin{equation}
r_{\rm h}^4 \Lambda+5 \Lambda l^2 r_{\rm h}^2+15(\lambda -12)=0.
\label{eq:large_r}
\end{equation}
It is clear that for  $\lambda \geq 12$, the above equation has no positive root which indicates absence of cosmological horizon. This means cosmological horizon can only occur when $\lambda < 12$.

On the other hand, for small $r$ limit, it is
\begin{equation}
15 r_{\rm h}^2(\lambda -12)+60 M r_{\rm h}-5 l^2(\lambda -36)=0,
\label{eq:small_r}
\end{equation}
which indicates that for $\lambda <12$ or $\lambda >36$, there always occurs one positive root giving an event horizon. That is, for event horizon to occur either $\lambda <12$ or $\lambda >36$. However referring to the full \ref{eq:Xr}, it becomes evident that there can occur no horizon for the window,  $12 \leq \lambda \leq 36$ as it can admit no positive roots.

This simple analysis has clearly shown the horizon structure of spacetime. Combining the two limits, large and small $r$ together, it follows that both event and cosmological horizons can occur for $\lambda <12$ and only event horizon for $\lambda >36$ and none for outside this prescription. We shall do a more detailed analysis in the next section.

\section{Physical viability of the solution} \label{sec:structure}
It turns out that pure GB solution with product topology, as in the solution in \ref{eq:EH_GB_S2 X S2_New} \cite{Dadhich:2015nua,Boulware:1985wk}, cannot be valid for all range of parameters for the discriminat, $h(r) \geq0$, which in turn demands presence of positive $\Lambda >0$. Another feature is that $h(r)=0$ gives a non-central singularity (there however occurs no central singularity when NUT charge is present) which has to lie inside the event horizon for physical viability of the solution. These two requirements are in general shared by all pure GB BHs with product topology including the present one. They strongly constrain the parameter space for  physical viability of spacetime in question.

Our aim is to find the parameter window for which $h(r) >0$ always outside the event horizon. This would ensure regularity of  spacetime in the physically accessible region lying between event and cosmological horizon. The two curves in \ref{Figure_2} to \ref{Figure_5} respectively refer to plots of $h(r)$, black and $\mathcal{X}(r)$, red. The intersection of these curves with the $x$-axis (which represents the radial distance $r$) denote location of singularity and horizon respectively. These plots capture all the possible scenarios, which may or may not be physically viable. In order to have an orderly study, we may divide the entire spectrum in different parts based on value of the parameter, $\lambda=\Lambda l^4$. There are four distinct possibilities: (a) $\lambda<12$, (b) $12 \leq \lambda<24$, (c) $24 \leq \lambda \leq 36$ and (d) $\lambda>36$ which we now take up.
\subsection{$\mathbf{\lambda<12}$}
\begin{figure}[htp]
\subfloat[In the above figure, there exist both the horizons, event and cosmological for $l=1.56M$. There are also three singularities, and two of which are naked, as they are not covered by the event horizon. Therefore, this option is not physically viable and hence ruled out.\label{Figure_02A}]{%
  \includegraphics[height=5.cm,width=.49\linewidth]{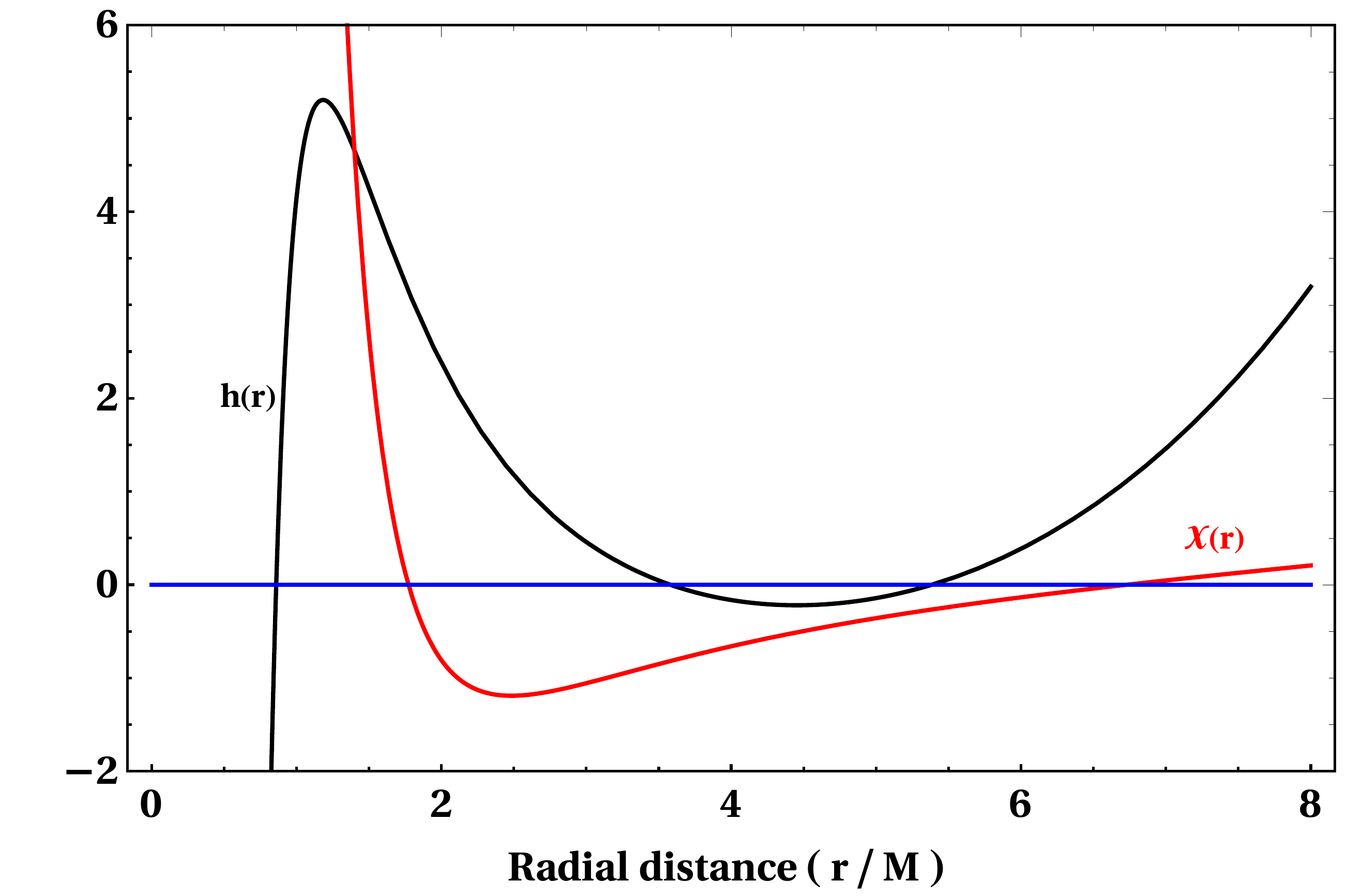}%
}\hfill
\subfloat[This is a physically viable case, as the singularity is covered by the event horizon and hence not naked. In addition to the event horizon, the cosmological horizon is also present. 
The NUT charge is fixed at $l=1.82M$.\label{Figure_02B}]{%
  \includegraphics[height=5.cm,width=.49\linewidth]{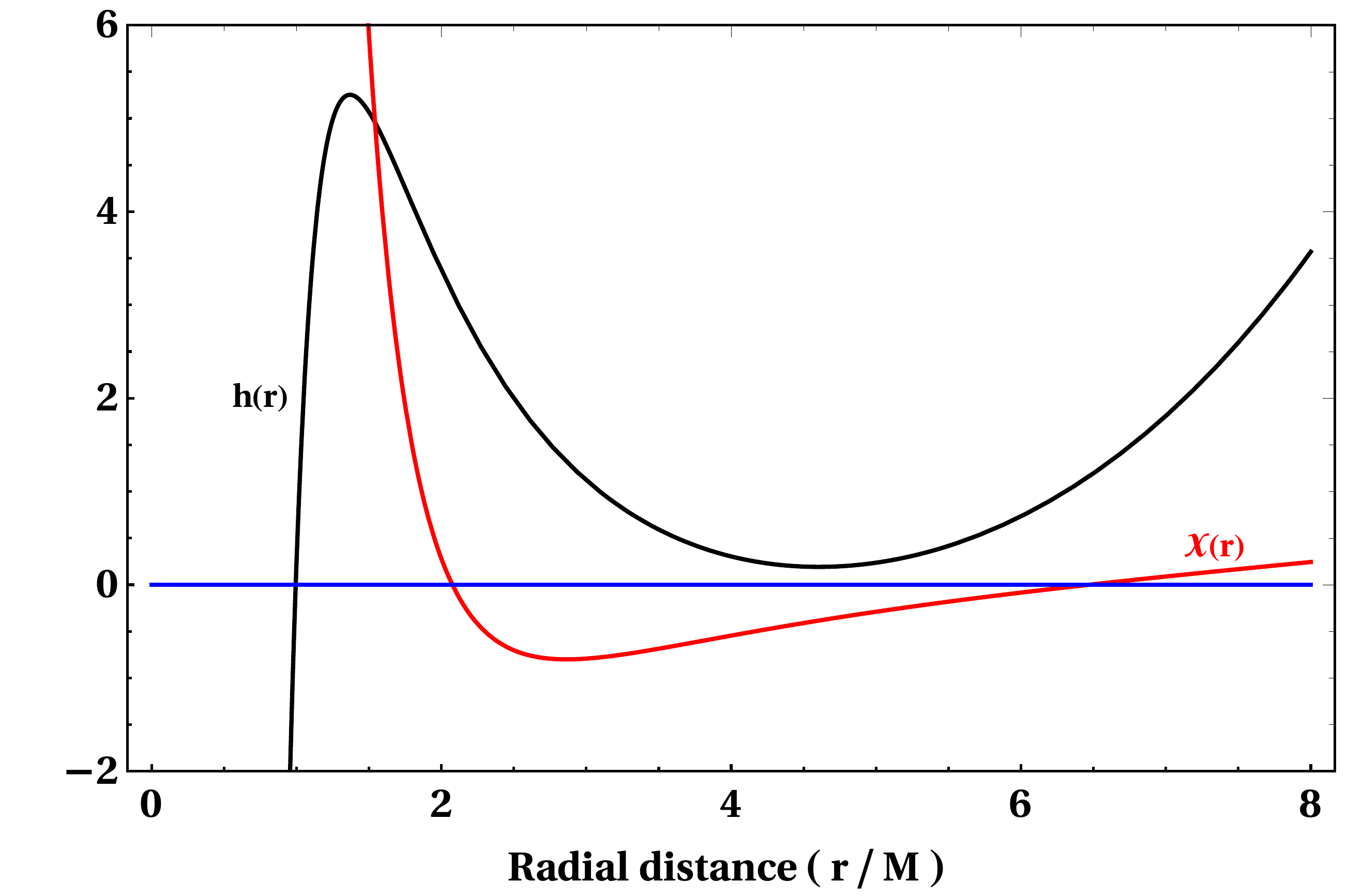}%
}\\
\hfill
\centering
\captionsetup{width=1\linewidth}
\subfloat[This is a non-viable possibility where a naked singularity is present, as there is no horizon. To reproduce the above plot, we choose $l=3.5M$.\label{Figure_02C}]{%
  \includegraphics[height=5.cm,width=.49\linewidth]{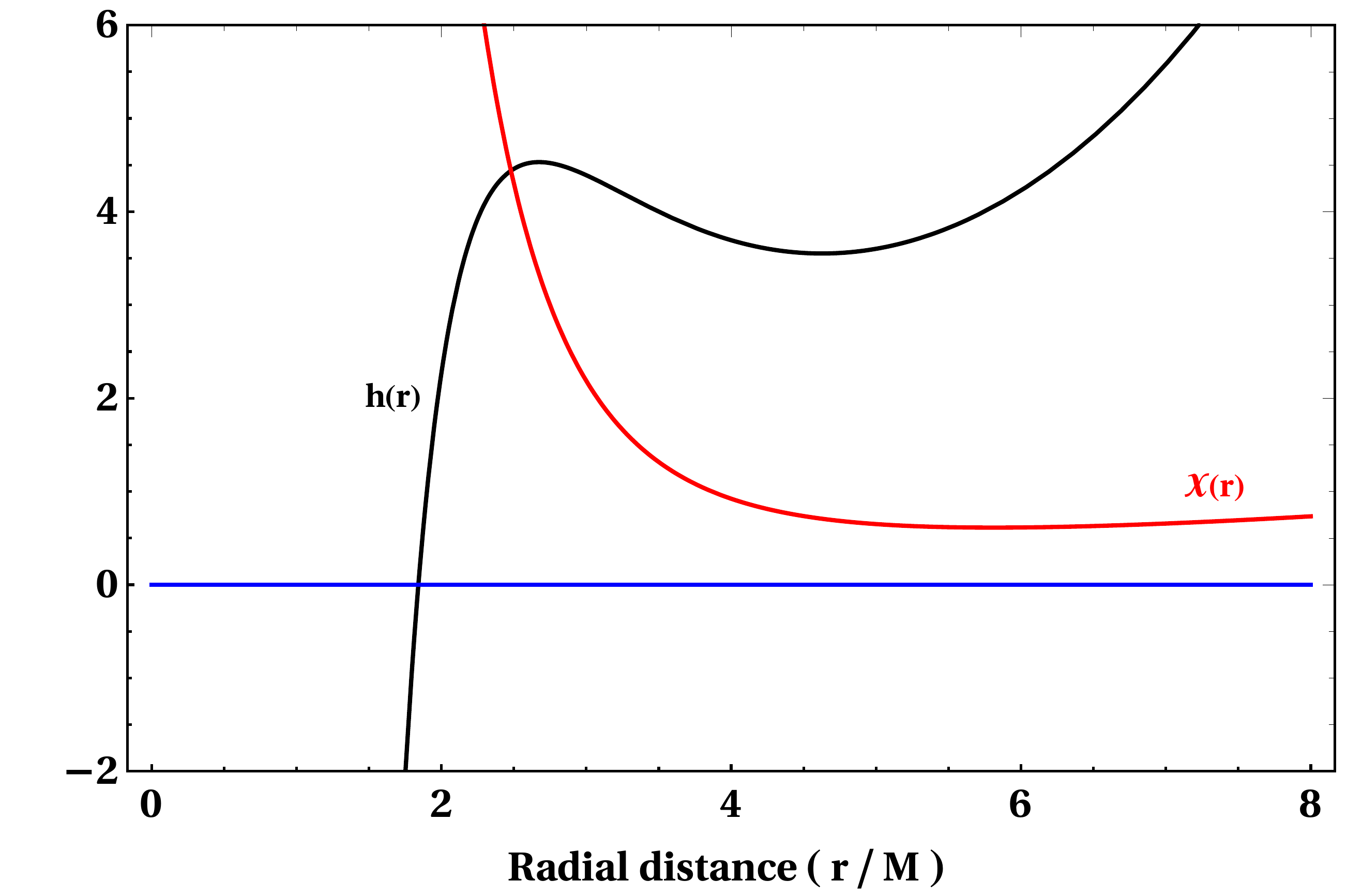}%
}
\caption{The above figures demonstrate three possible cases with $\lambda <12$, and each of the figures has $\Lambda M^4=0.06$.}
\label{Figure_2}
\end{figure}

In this case, \ref{C_1_0_C_2_0_NEW} can either have 1 or 3 positive root(s), which implies presence of at least one singularity always. Besides, the horizon equation given by \ref{eq:Xr} can have either 0 or 2 positive roots. Therefore, in principle, there can be four possibilities such as --- (i) three singularities, two horizons, (ii) one singularity, two horizons and the singularity is hidden behind the event horizon, (iii) one singularity and no horizon and (iv) three singularities and no horizon. We straightway rule out the last case. Out of the other three cases, as depicted in \ref{Figure_2}, only the second case is physically viable in which the singularity is covered by the event horizon. For $\Lambda M^4=0.06$ and $\left\{12/\Lambda\right\}^{1/4}=3.76M$, the first option is realized only for $0<l \lessapprox 1.70M$; while for second and third cases it is $1.70M \lessapprox l \lessapprox 2.64M$ and $2.64M \lessapprox l < 3.76M$ respectively. 
\subsection{$\mathbf{12 \leq \lambda \leq 24}$}
\begin{figure}[htp]
\centering
\includegraphics[scale=.28]{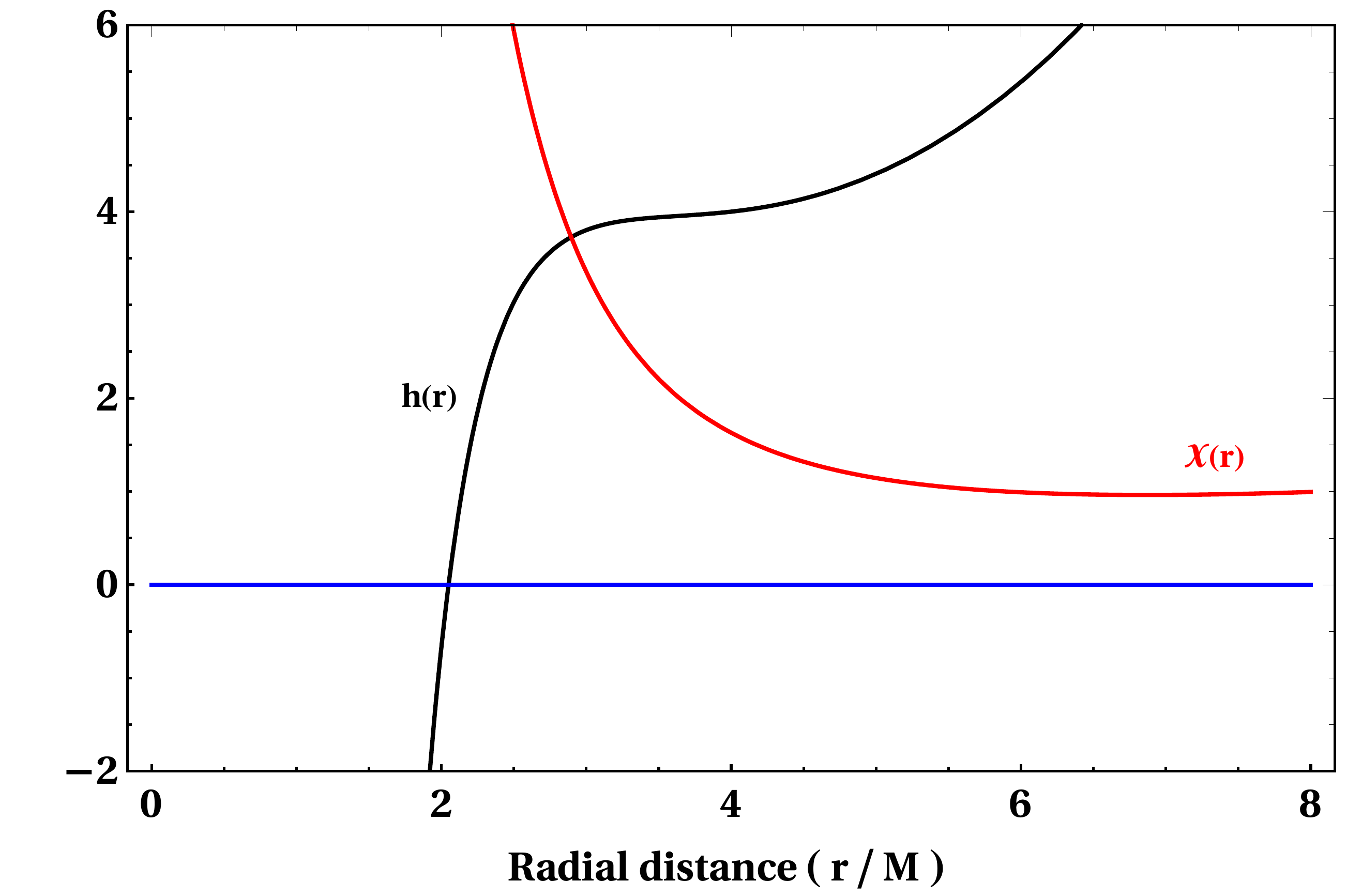}
\caption{It is shown that the singularity is always naked and there is neither event nor cosmological horizon. Here we set $\Lambda M^4= 0.06$ and $l=4M$. Given that the singularity is naked, the spacetime is not physically viable.}
\label{Figure_3}
\end{figure}

Within this range, the singularity equation always have one positive root, while the horizon equation has no positive root. This would indicate that singularity will always remain naked, hence this is not physically acceptable. This is shown in \ref{Figure_3}, where we take $\Lambda M^4=0.06$, and the bound on the NUT charge is given as $\left\{{12}/{\Lambda}\right\}^{1/4}\simeq 3.76M \leq l \leq \left\{{24}/{\Lambda}\right\}^{1/4} \simeq 4.47M$.
\subsection{$\mathbf{24 \leq \lambda \leq 36}$}
\begin{figure}[htp]
\subfloat[In this example, the spacetime has two singularities which are naked and hence it is physically ruled out. The value of the NUT charge is fixed at $l=4.5M$.\label{Figure_04A}]{%
  \includegraphics[height=5.5cm,width=.49\linewidth]{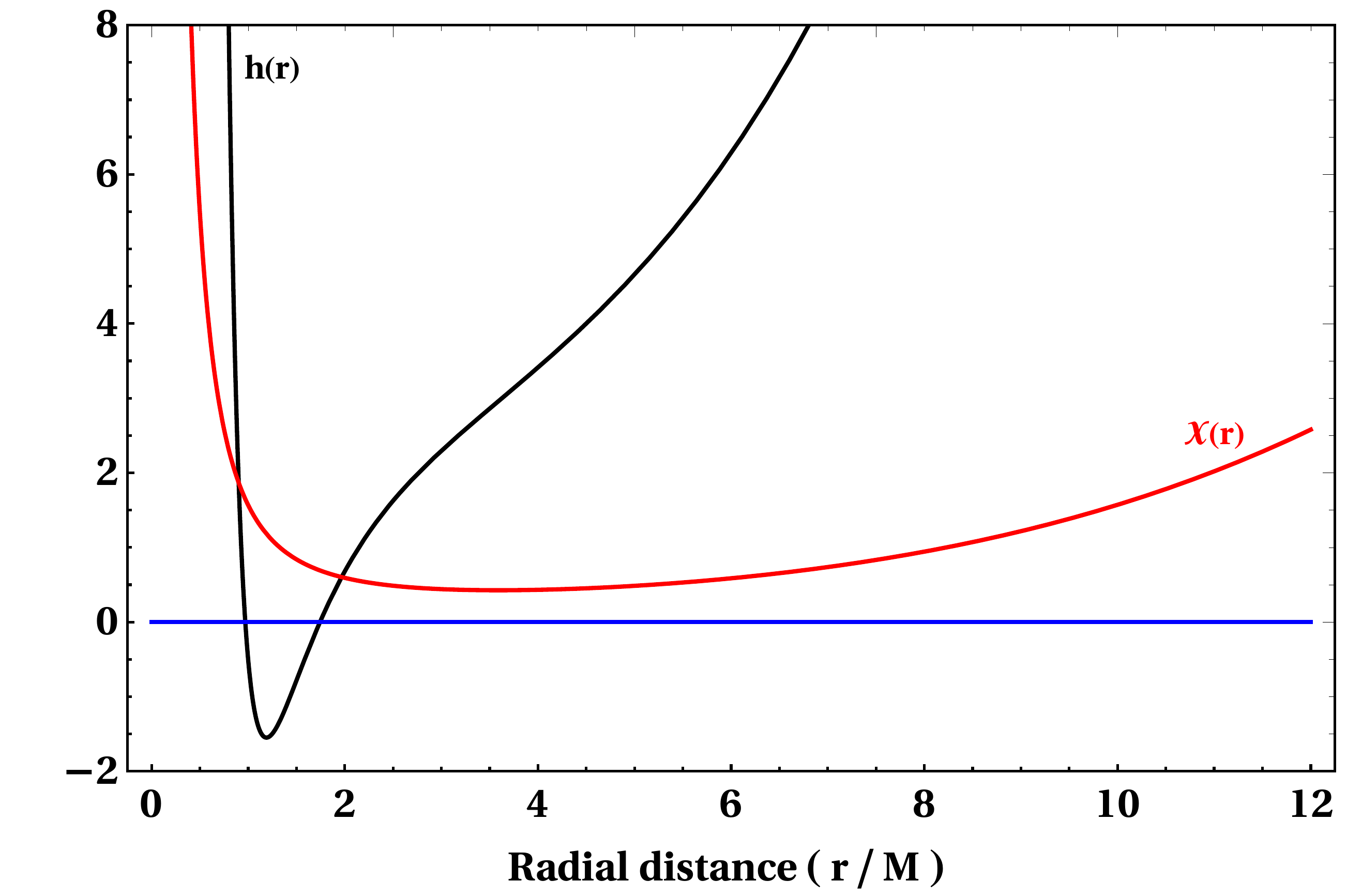}%
}\hfill
\subfloat[In this case, there occur neither singularity not horizons and hence it presents an interesting possibility where the spacetime is regular everywhere. Here we have set $l=4.9M$.\label{Figure_04B}]{%
  \includegraphics[height=5.5cm,width=.49\linewidth]{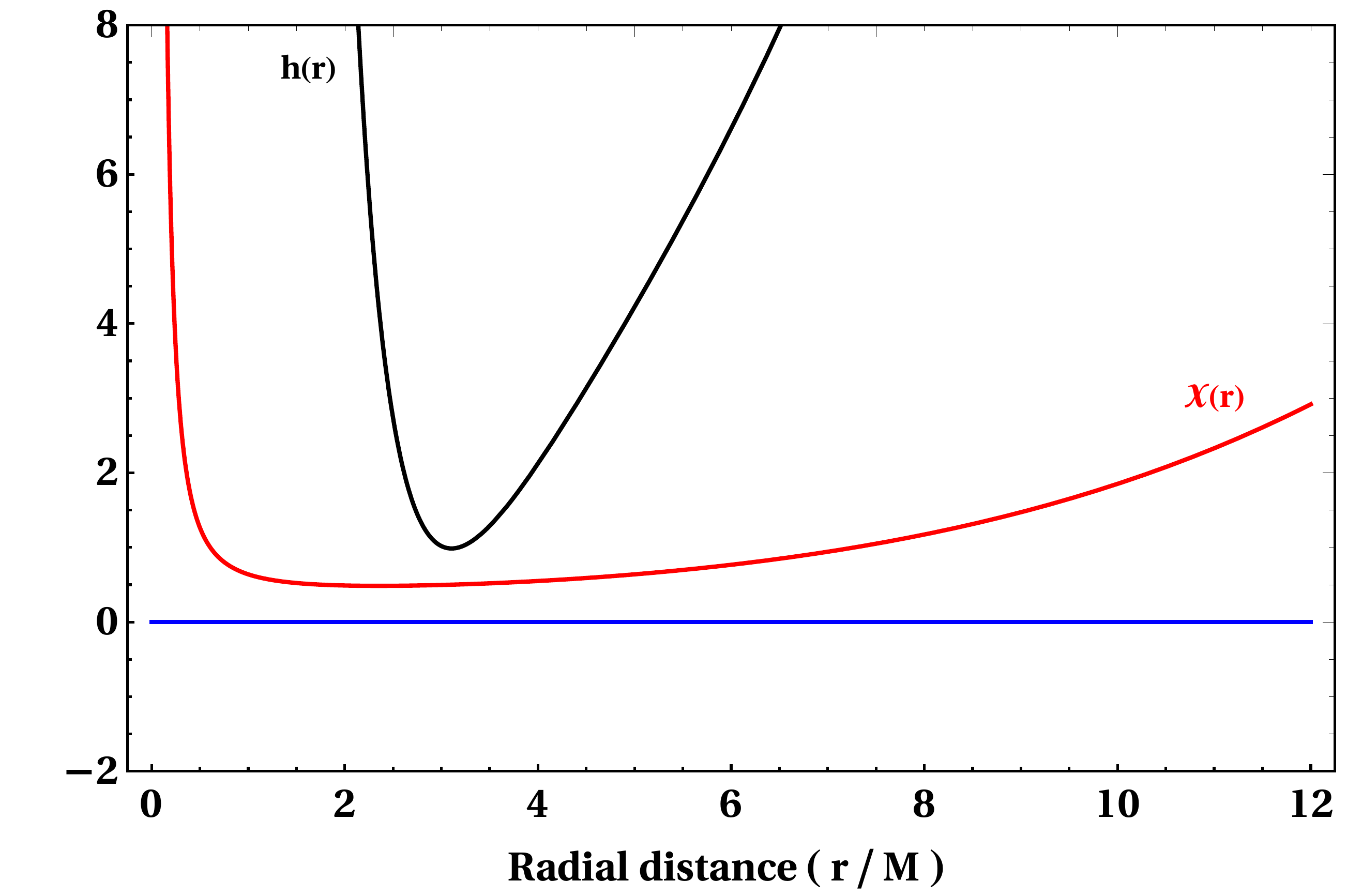}%
}
\caption{The above figure illustrate both the possibilities for $24 \leq \lambda \leq 36$ where we have set $\Lambda M^4=0.06$.}
\label{Figure_4}
\end{figure}

For this range, \ref{eq:Xr} has no positive root, which indicates that there can occur neither event nor cosmological horizon. However \ref{C_1_0_C_2_0_NEW} can have 0 or 2 positive roots. That means either spacetime is free of singularity or if they occur, they would be naked. The former possibility presents an interesting case of regular spacetime free of singularity as well as  horizons. Both of these possibilities are shown in \ref{Figure_4}. Similar to the earlier cases, here also the value of NUT parameter is severely constrained. For example, with $\Lambda M^4=0.06$, if singularity is naked, then the NUT charge is constrained to  $\left\{24/\Lambda\right\}^{1/4} \simeq 4.47 \leq l \lessapprox 4.51M$. On the other hand, if the spacetime is free of both singularity and horizons, it is bound as $4.51M \lessapprox l \leq \left\{36/\Lambda\right\}^{1/4} \simeq 4.95M$.
\subsection{$\mathbf{\lambda>36}$}
\begin{figure}[htp]
\subfloat[In the above figure, the spacetime is free of singularity and contains the event horizon, and therefore, the spacetime is physically viable. The NUT charge is fixed at $l=4.96M$.\label{Figure_05A}]{%
  \includegraphics[height=5.5cm,width=.49\linewidth]{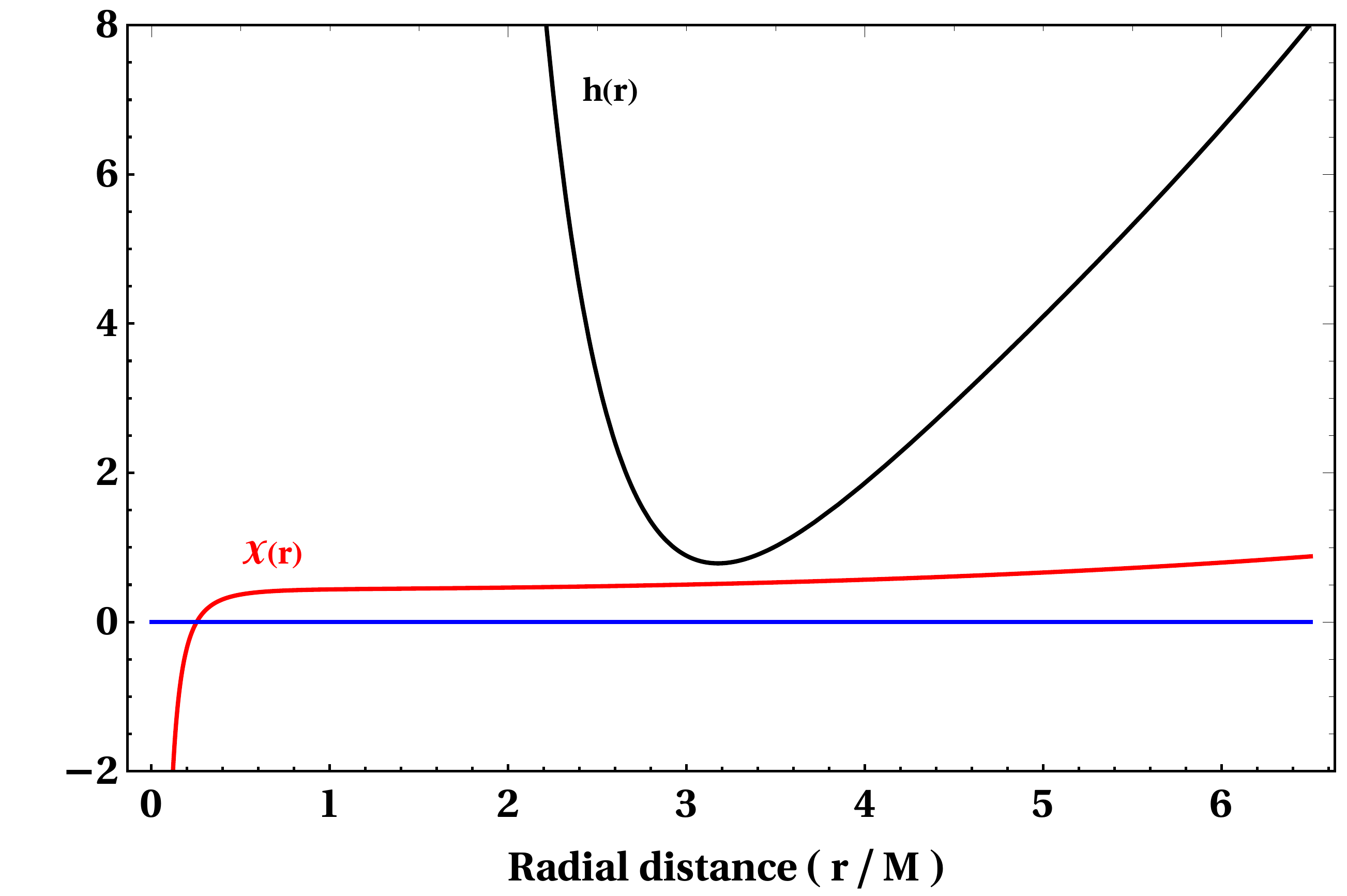}%
}\hfill
\subfloat[In this case, both the singularities are naked and hence physically unacceptable. We fix the NUT charge at $l=6M$.\label{Figure_05B}]{%
  \includegraphics[height=5.5cm,width=.49\linewidth]{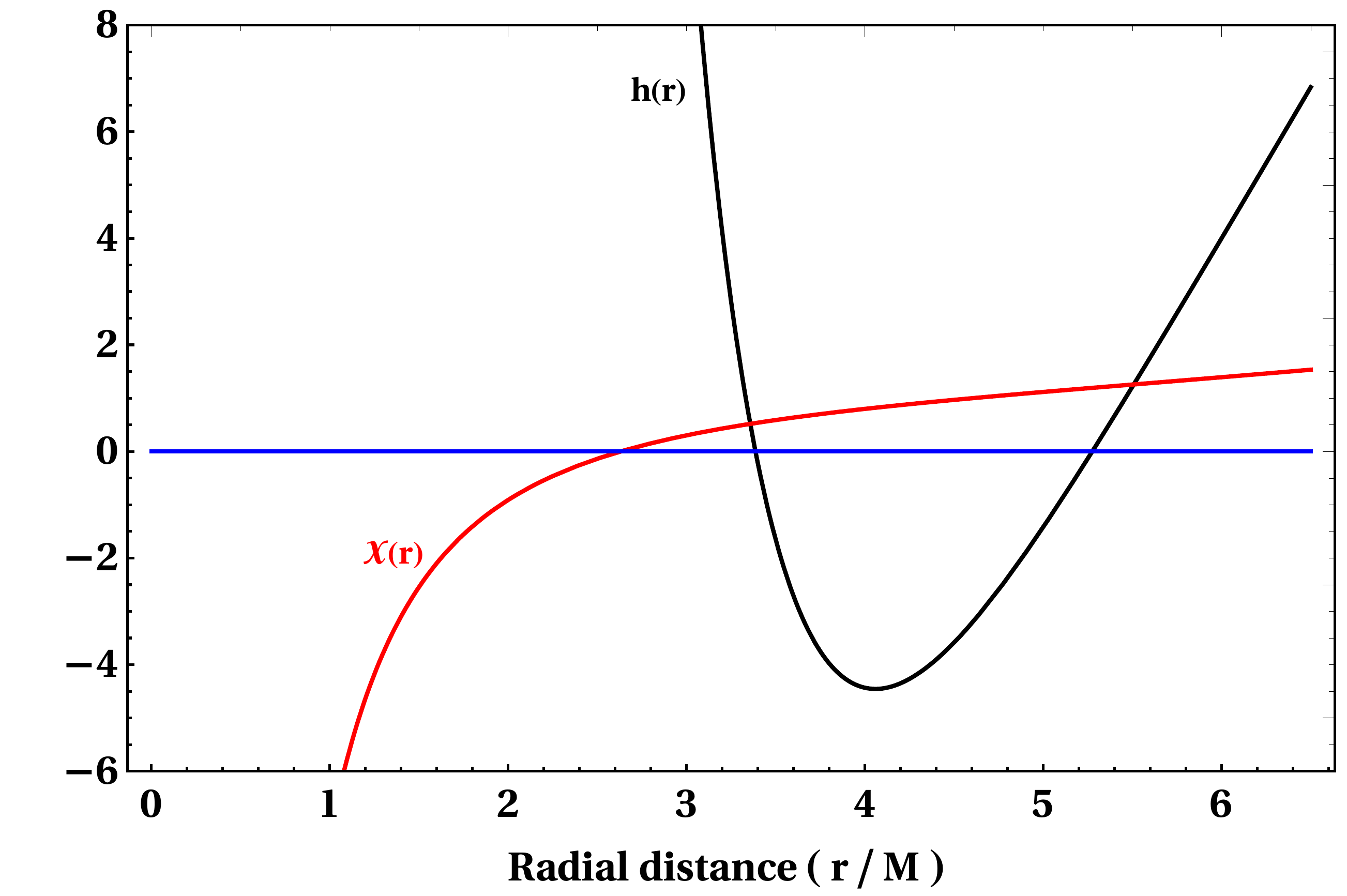}%
}
\caption{The above figures illustrate the two examples with $\lambda>36$, where we fix the cosmological constant $\Lambda M^4=0.06$.}
\label{Figure_5}
\end{figure}

In this case, there always occurs one horizon while there may either occur none or two singularities. By referring to  \ref{eq:large_r} and \ref{eq:small_r}, it becomes evident that this horizon can only be event and not cosmological. Out of these two cases, the first one is only viable because in the other both the singularities are naked. These two cases are shown in \ref{Figure_5} for $\Lambda M^4=0.06$. For the first option, the NUT charge is constrained as $\left\{36/\Lambda\right\}^{1/4} \simeq 4.95M \leq l \lessapprox 5.17 M$ while for the second, $5.17M \lessapprox l < \infty$.
\section{Discussion and conclusion} \label{sec:discussion}
In this paper we have found an exact solution of pure GB $\Lambda$-vacuum equation describing a BH with NUT charge. Its horizon has product, $S^{(2)} \times S^{(2)}$ topology. This setting of pure GB and product topology has characteristic properties. One, $\Lambda >0$ is required for large $r$ validity of the solution and two, occurrence of non-central singularity when discriminant vanishes. A suitable parameter window has to be found for which either non-central singularity does not occur; i.e., discriminant $>0$ or if it does, it is hidden behind event horizon \cite{Dadhich:2015nua,Boulware:1985wk}. This is so even in absence of NUT parameter \cite{Pons:2014oya}. NUT charge brings in its own further complexities. It however removes central singularity, but structure of non-central singularity and horizons becomes highly involved.

Thus occurrence of non-central singularity is a generic feature of pure GB BH with product topology. The most troublesome part is how to avoid naked singularity given by \ref{C_1_0_C_2_0_NEW}, $h(r) =0$. This is a polynomial equation which may have number of positive roots which have to be avoided or to be covered by event horizon so that physically accessible region is good and regular.

Another point of concern is the horizon equation, \ref{eq:Xr}, which has either one, two or none positive roots. Since $\Lambda >0$ is present, there should occur both event and cosmological horizons. That happens only for the dimensionless parameter, $\lambda = \Lambda l^4 < 12$. It turns out that for this range, there occurs only one singularity which could indeed be hidden behind the event horizon as shown in \ref{Figure_02B}. This is the most acceptable BH spacetime with expected behaviour of singularity being covered by the event horizon so that physically accessible region bounded by event and cosmological horizons is benign and regular.

On the other hand for $\lambda > 36$, there occurs only event horizon, and neither singularity nor cosmological horizon occurs as depicted in \ref{Figure_05A}. This is rather unusual because $\Lambda >0$ should generally give cosmological horizon.

It turns out that there can occur no horizon for the entire range,  $12 \leq l \leq 36$ while in its sub range $24 \leq l \leq 36$, there is a possibility of having no singularity either. This then presents a remarkably intriguing situation of spacetime being free of both horizon as well as singularity;i.e. regular everywhere for  $0 \leq r \leq \infty$. This is all very fine except one doesn't understand where and how does its source reside? Is it cosmological or describes a compact object? It therefore calls for further investigation for clearer understanding which we intend to take up in future.

It is only the case of $l < 12$ which describes a NUT BH with both event and cosmological horizons and the region enclosed by
them is regular, free of singularity. In the other two cases, one of all through regular spacetime without horizon and the other of BH having only the event but no cosmological horizon, the situation is rather unusual and hence further study is called for their overall physical understanding. However it is remarkable that despite there being sever restrictions imposed by singularity and horizon equations \ref{C_1_0_C_2_0_NEW} and \ref{eq:Xr} respectively, yet there exists a parameter window giving a well behaved and physically viable BH spacetime.

In the above discussion, it is the dimensionless parameter, $\lambda = \Lambda l^4$, which plays the determining role in occurrence of horizon and singularity. In contrast to the BH charge parameters, $M$ and $l$, it is curiously interesting  that this hybrid parameter arising from clubbing of $l$ and $\Lambda$ determines the overall physical character of spacetime. For $\lambda <12$, both event and cosmological horizons occur, then they both disappear in the intermediate range, $12 \leq \lambda \leq 36$ and finally for $\lambda >36$ only event horizon reappears. It is the combined effect of $l$, $\Lambda$ and product topology in pure Gauss-Bonnet framework. It is rather hard to disentangle various effects, all that happens is that the hybrid dimensionless parameter $\lambda$ acquires the determining role.

On the way to the solution, we discuss how topology plays a key role in understanding higher dimensional NUT BHs. More precisely, we attempt to answer the question why is it that all known NUT solutions, see \cite{Flores-Alfonso:2018jra,Awad:2000gg}, in higher dimensions always have product topology $S^{(2)} \times S^{(2)}$? In the present example of a six-dimensional NUT spacetime, we found that $S^{(2)}\times S^{(2)}$ topology makes the Riemann components independent of the angular coordinate, while $S^{(4)}$ does not. Similarly, for the $S^{(1)} \times S^{(3)}$ horizon topology, we found that Riemann components are not free of $\theta$ part. We have further verified that for the eight dimensional case, the product topology of $S^{(3)} \times S^{(3)}$ is not compatible with the property of Riemann being free of $\theta$ coordinate. This leads us to speculate that the very existence (or non-existence) of a higher dimensional NUT solution, may critically depend on the fact whether Riemann are independent of $\theta$ or not. For example, in the present case of six dimensional, the NUT vacuum solution exists for $S^{(2)}\times S^{(2)}$ and not for $S^{(4)}$. However, we believe a detailed analysis is required to establish this claim on more robust and firmer ground. This is however beyond the scope of this paper. If it is the case that only $S^{(2)}$ or its product is allowed for the horizon topology, there could exist only  the even dimensional NUT BHs in $D>4$. We intend to take up this question in future studies.

Finally we end up with the important question that has emerged in this investigation, how does compatibility of radial dependence of the curvature picks out $S^{(2)}$ or its product topology for NUT BH horizon? It may be noted that unlike mass and electric charge, which are extensive physical parameters, NUT parameter like rotation also participates in defining spatial geometry. The former two appear only as source in potential while rotation and NUT parameters also define space geometry through $r^2 + a^2 \cos^2\theta$ and $r^2 + l^2$ respectively. But for  the latter, $r^2 + l^2$ is function of $r$ alone which may be the reason for Riemann being function of $r$ only. Yet it has perhaps to be probed whether could there exist NUT vacuum solution with Riemann curvature being function of $r$ and $\theta$? Another related question is of characterising this  property rigorously in terms of Killing symmetry and groups of motion. We wish to take up these questions in future.

%
\section*{Acknowledgement}
The authors would like to thank Sumanta Chakraborty for useful discussions. S.M. wishes to thank Department of Science and Technology (DST), Government of India, and Lumina Quaeruntur No. LQ100032102 of the Czech Academy of Sciences, for financial support. N.D. acknowledges 
 the support of the CAS President's International  Fellowship Initiative Grant No. 2020VMA0014.
\bibliography{References}
\bibliographystyle{./utphys1}
\end{document}